\documentclass[twocolumn,preprintnumbers,amsmath,amssymbm,prd]{revtex4}
\usepackage{epsfig}
\usepackage{graphicx}

\begin{document}
\title{Analytic treatment of the charged black-hole-mirror bomb in the highly explosive regime}
\author{Shahar Hod}
\affiliation{The Ruppin Academic Center, Emeq Hefer 40250, Israel}
\affiliation{ } \affiliation{The Hadassah Institute, Jerusalem
91010, Israel}
\date{\today}

\begin{abstract}
\ \ \ A charged scalar field impinging upon a charged
Reissner-Nordstr\"om black hole can be amplified as it scatters off
the hole, a phenomenon known as superradiant scattering. This
scattering process in the superradiant regime $\omega<qQ/r_+$ (here
$\omega, q, Q$, and $r_{\pm}$ are the conserved frequency of the
wave, the charge coupling constant of the field, the electric charge
of the black hole, and the horizon radii of the black hole,
respectively) results in the extraction of Coulomb energy and
electric charge from the charged black hole. The black-hole-field
system can be made unstable by placing a reflecting mirror around
the black hole which prevents the amplified field from escaping to
infinity. This charged black-hole-mirror system is the spherically
symmetric analogue of the rotating black-hole-mirror bomb of Press
and Teukolsky. In the present paper we study {\it analytically} the
charged black-hole-mirror bomb in the asymptotic regime $qQ\gg1$ and
for mirror radii $r_{\text{m}}$ in the near-horizon region
$x_{\text{m}}\equiv(r_{\text{m}}-r_+)/r_+\ll\tau$, where $\tau\equiv
(r_+-r_-)/r_+$ is the dimensionless temperature of the black hole.
In particular, we derive analytic expressions for the oscillation
frequencies $\Re\omega$ and the instability growth timescales
$1/\Im\omega$ of the superradiant confined fields. Remarkably, we
find a simple linear scaling $\Im\omega\propto qQ/r_+$ for the
imaginary part of the resonances in the asymptotic $qQ\gg
(\tau/x_{\text{m}})^2\gg1$ regime, which implies that the
instability timescale $1/\Im\omega$ of the system can be made
arbitrarily short in the $qQ\to\infty$ limit. The short instability
timescale found in the linear regime along with the spherical
symmetry of the system, make the charged bomb a convenient toy model
for future numerical studies aimed to investigate the non-linear
end-state of superradiant instabilities.
\end{abstract}
\bigskip
\maketitle

%]

\section{Introduction}

Black holes are believed to be the most powerful source of energy in
the Universe. The mass-energy $M$ of a rotating Kerr black hole of
angular-momentum $J$ can be expressed in the form
\cite{Ch,ChRu,Noteun}
\begin{equation}\label{Eq1}
M=\sqrt{M^2_{\text{ir}}+{{J^2}\over{4M^2_{\text{ir}}}}}\  ,
\end{equation}
where $M_{\text{ir}}$ is the irreducible mass of the black hole
which is closely related to its surface area $A$:
%\begin{equation}\label{Eq1}
$M_{\text{ir}}=\sqrt{{{A}/{16\pi}}}$.
%\end{equation}

The well-known area theorem of Hawking \cite{Haw} reveals that,
within the framework of classical general relativity, the black-hole
surface area (and thus also the irreducible mass) cannot decrease. A
physical process in which the irreducible mass of the black hole
remains unchanged is known as a reversible transformation
\cite{Ch,ChRu}. The relation $M=M(J,M_{\text{ir}})$, Eq.
(\ref{Eq1}), implies that up to $\sim29\%$ (a fraction
$1-1/\sqrt{2}$) of the energy of a Kerr black hole is in the form of
rotational energy which, in principle, can be released in a
reversible process \cite{Noteext} (in such a process the black-hole
parameters are changed according to $M\to M_{\text{ir}}$ and $J\to
0$).

One possible mechanism to extract the rotational energy of a Kerr
black hole is based on the well-known phenomenon of superradiant
scattering \cite{Zel,Viln}: a bosonic field of the form
$e^{im\phi}e^{-i\omega t}$ can be amplified as it scatters off a
rotating Kerr black hole of angular-velocity $\Omega_{\text{H}}$ if
it respects the superradiance bound \cite{Zel}
\begin{equation}\label{Eq2}
\omega<m\Omega_{\text{H}}\  .
\end{equation}
The amplification of the incident bosonic field in the superradiant
regime (\ref{Eq2}) signals a decrease in the rotational energy of
the black hole \cite{Zel}.

Press and Teukolsky \cite{PressTeu2} have pointed out that the
mechanism of superradiant scattering can be used in order to build a
powerful bomb whose energy source is the rotational energy of the
black hole itself. To build this {\it black-hole bomb} one should
prevent the amplified scattered field from escaping to infinity. The
original suggestion of Press and Teukolsky was to surround the black
hole by a reflecting mirror \cite{PressTeu2,Notemas}. In this way
the bosonic field [a wave packet made of frequencies in the
superradiant regime (\ref{Eq2})] will bounce back and forth between
the black hole and the mirror amplifying itself each time. As a
consequence, the rotational energy extracted from the black hole by
the trapped bosonic field would grow exponentially over time
\cite{PressTeu2}. Using numerical techniques, it was found in
\cite{CarDias} that the maximum growth rate of a scalar field (the
largest imaginary part of the superradiant resonance frequency) in
the Kerr black-hole-mirror system is given by
\begin{equation}\label{Eq3}
\Im\omega\simeq 6\times 10^{-5}M^{-1}\  .
\end{equation}

An analogous superradiant amplification of waves may take place when
a {\it charged} bosonic field impinges upon a {\it charged}
Reissner-Nordstr\"om (RN) black hole \cite{Bekch}. In the charged
case the superradiant scattering (amplification of the waves) occurs
for incident waves with frequencies in the regime \cite{Bekch}
\begin{equation}\label{Eq4}
\omega<q\Phi_{\text{H}}\  ,
\end{equation}
where $q$ is the charge coupling constant of the field and
\begin{equation}\label{Eq5}
\Phi_{\text{H}}={{Q}\over{r_+}}\
\end{equation}
is the electric potential of the RN black hole. [Here $Q$ and $r_+$
are the electric charge and horizon radius of the black hole,
respectively].

The scattering of charged scalar fields off a charged RN black hole
in the superradiant regime (\ref{Eq4}) results in the extraction of
Coulomb energy and electric charge from the charged black hole
\cite{Bekch}. The mass-energy $M$ of a RN black hole of charge $Q$
can be expressed in the form \cite{Ch,ChRu,Noteun}
\begin{equation}\label{Eq6}
M=M_{\text{ir}}+{{Q^2}\over{4M_{\text{ir}}}}\  .
\end{equation}
The relation $M=M(Q,M_{\text{ir}})$, Eq. (\ref{Eq6}), implies that
up to $50\%(!)$ of the energy of a charged RN black hole is in the
form of a Coulomb energy which, in principle, can be released in a
reversible process \cite{Noteext2} (in such a process the black-hole
parameters are changed according to $M\to M_{\text{ir}}$ and $Q\to
0$).

The physical interest in RN black holes is mainly motivated by the
fact that these charged black holes share many common
characteristics with the astrophysically more relevant Kerr black
holes. In particular, the global spacetime structures of charged RN
black holes and rotating Kerr black holes are almost identical
\cite{HodPirmass}. It is therefore of physical interest to explore
the properties of the {\it charged} black-hole-mirror bomb, which is
the spherically symmetric analogue of the {\it rotating}
black-hole-mirror bomb of Press and Teukolsky \cite{PressTeu2}.

In addition, the fact that the charged black-hole-mirror bomb has
spherical symmetry (as opposed to the non-spherically symmetric Kerr
black-hole spacetime) makes it a convenient toy model for future
numerical studies aimed to investigate the non-linear dynamics of
explosive superradiant instabilities. The present analytical study,
which is restricted to the linear regime, should be regarded as a
first step in this direction.

In a very interesting work, Degollado et. al. \cite{Dego} have
recently studied this charged black-hole-mirror system (still
restricted to the linear level) using numerical techniques.
Remarkably, the authors of \cite{Dego} reported on instability
growth rates of the superradiant charged confined fields which are
several orders of magnitude {\it larger} than the maximal growth
rate (\ref{Eq3}) found for the rotating black-hole-mirror bomb.

The numerical results presented in \cite{Dego} indicate that
$\Im\omega$ increases monotonically with increasing values of the
charge coupling constant $q$ of the field. Unfortunately, the
authors of \cite{Dego} also stated that their numerical scheme
breaks down for large values of the parameter $q$. For this reason,
the largest imaginary part of the superradiant resonance frequency
reported in \cite{Dego} is
\begin{equation}\label{Eq7}
\Im\omega\sim 0.07M^{-1}\  .
\end{equation}
As emphasized in \cite{Dego}, the reported value (\ref{Eq7}) is not
the maximum possible value of $\Im\omega$. It is merely the maximal
value of $\Im\omega$ which could be obtained numerically under the
technical limitations imposed by the numerical tools used in
\cite{Dego}.

The main goal of the present paper is to explore the physical
properties of the charged black-hole-mirror bomb using {\it
analytical} techniques. As we shall show below, the instability
growth rate of the superradiant confined charged fields (the value
of $\Im\omega$) can grow unboundedly in the $qQ\to\infty$ limit.

\section{Description of the system}

The physical system we consider consists of a charged scalar field
$\Psi$ linearly coupled to a charged RN black hole. The black-hole
spacetime is described by the line element
\begin{equation}\label{Eq8}
ds^2=-f(r)dt^2+{1\over{f(r)}}dr^2+r^2(d\theta^2+\sin^2\theta
d\phi^2)\ ,
\end{equation}
where
\begin{equation}\label{Eq9}
f(r)\equiv 1-{{2M}\over{r}}+{{Q^2}\over{r^2}}\  .
\end{equation}
Here $M$ and $Q$ are respectively the mass and electric charge of
the black hole, and $r$ is the Schwarzschild areal coordinate. The
zeros of $f(r)$,
\begin{equation}\label{Eq10}
r_{\pm}=M\pm (M^2-Q^2)^{1/2}\  ,
\end{equation}
are the black-hole (event and inner) horizons.

The dynamics of the charged scalar field $\Psi$ in the charged RN
spacetime is governed by the Klein-Gordon wave equation
\cite{HodPirpam,Stro,HodCQG2}
\begin{equation}\label{Eq11}
[(\nabla^\nu-iqA^\nu)(\nabla_{\nu}-iqA_{\nu}) -\mu^2]\Psi=0\  ,
\end{equation}
where $A_{\nu}=-\delta_{\nu}^{0}{Q/r}$ is the electromagnetic
potential of the black hole. Here $q$ and $\mu$ are respectively the
charge and mass of the field \cite{Noteuni}. One may decompose the
field $\Psi$ in the form
\begin{equation}\label{Eq12}
\Psi_{lm}(t,r,\theta,\phi)=e^{im\phi}S_{lm}(\theta)R_{lm}(r)e^{-i\omega
t}\ ,
\end{equation}
where $\omega$ is the conserved frequency of the mode and $\{l,m\}$
are respectively the spherical harmonic index and the azimuthal
harmonic index of the mode (we shall henceforth omit the indices $l$
and $m$ for brevity). The sign of $\Im\omega$ determines whether the
solution is stable (decaying in time with $\Im\omega<0$) or unstable
(growing in time with $\Im\omega>0$). Stationary modes are
characterized by $\Im\omega=0$.

Substituting the decomposition (\ref{Eq12}) into the Klein-Gordon
wave equation (\ref{Eq11}), one finds \cite{HodPirpam,Stro,HodCQG2}
that $R(r)$ and $S(\theta)$ obey radial and angular equations both
of confluent Heun type \cite{Heun,Abram} coupled by a separation
constant $K_l=l(l+1)$, where $l\geq |m|$ is an integer. The radial
wave equation is given by \cite{HodPirpam,Stro,HodCQG2}
\begin{equation}\label{Eq13}
\Delta{{d} \over{dr}}\Big(\Delta{{dR}\over{dr}}\Big)+UR=0\ ,
\end{equation}
where
\begin{equation}\label{Eq14}
\Delta\equiv r^2-2Mr+Q^2\  ,
\end{equation}
and
\begin{equation}\label{Eq15}
U\equiv(\omega r^2-qQr)^2 -\Delta[\mu^2r^2+l(l+1)]\  .
\end{equation}

We are interested in solutions of the radial equation (\ref{Eq13})
with the physical boundary conditions of purely ingoing waves at the
black-hole horizon and a vanishing field at the location
$r_{\text{m}}$ of the mirror \cite{CarDias,Dego}. That is,
\begin{equation}\label{Eq16}
%\label{eq:boundary_conditions}
R \sim e^{-i (\omega-qQ/r_+)y}\ \ \text{ as }\ r\rightarrow r_+\ \
(y\rightarrow -\infty)\ ,
%\begin{cases}
%{1\over r}e^{-\sqrt{\mu^2-\omega^2}y} & \text{ as }
%r\rightarrow\infty\ \ (y\rightarrow \infty)\ ; \\
%e^{-i (\omega-m\Omega)y} & \text{ as } r\rightarrow r_+\ \
%(y\rightarrow -\infty)\ ,
%\end{cases}
\end{equation}
and
\begin{equation}\label{Eq17}
R(r=r_{\text{m}})=0\  .
\end{equation}
Here the ``tortoise" radial coordinate $y$ is defined by
$dy=(r^2/\Delta)dr$. The boundary condition (\ref{Eq16}) describes
an outgoing flux of energy and charge from the charged black hole
for scattered fields in the superradiant regime (\ref{Eq4})
\cite{Bekch,Dego}.

The boundary conditions (\ref{Eq16})-(\ref{Eq17}) single out a
discrete set of complex resonances known as Boxed Quasi-Normal
frequencies $\{\omega^{\text{BQN}}_n(r_{\text{m}})\}$
\cite{CarDias,Noteqnm}. The main goal of the present paper is to
determine these characteristic resonances analytically in the
superradiant regime (\ref{Eq4}). To that end, it is convenient to
define new dimensionless variables
\begin{equation}\label{Eq18}
x\equiv {{r-r_+}\over {r_+}}\ \ ;\ \ \tau\equiv{{r_+-r_-}\over
{r_+}}\ \ ;\ \ \omega r_+\equiv qQ+\epsilon\  ,
%varpi\equiv{{2(\omega r_+ -qQ)}\over{\tau}}\ \ ;\ \ k\equiv 2\omega
%r_+-qQ\  ,
\end{equation}
in terms of which the radial wave equation (\ref{Eq13}) becomes
\begin{equation}\label{Eq19}
x(x+\tau){{d^2R}\over{dx^2}}+(2x+\tau){{dR}\over{dx}}+VR=0\  ,
\end{equation}
where
\begin{equation}\label{Eq20}
V\equiv K^2/x(x+\tau)-[\mu^2r^2_+(x+1)^2+l(l+1)]
\end{equation}
with
\begin{equation}\label{Eq21}
K\equiv (qQ+\epsilon)x^2+(qQ+2\epsilon)x+\epsilon\  .
\end{equation}

As we shall show below, the radial equation (\ref{Eq19}) is amenable
to an analytic treatment in the
%highly-charged-near-horizon
%near-horizon
%asymptotic regime
double limit
\begin{equation}\label{Eq22}
qQ\gg 1\ \ \ \text{with}\ \ \ x\ll\tau\  .
\end{equation}
In this asymptotic regime the radial equation (\ref{Eq19}) can be
approximated by
\begin{equation}\label{Eq23}
x^2{{d^2R}\over{dx^2}}+x{{dR}\over{dx}}+V_{\text{near}}R=0\  ,
\end{equation}
where \cite{NoteK}
\begin{equation}\label{Eq24}
V_{\text{near}}=\tau^{-2}[(qQ+2\epsilon)x+\epsilon]^2
%+O(x/\tau)
\ .
\end{equation}

\section{The stationary resonances of the black-hole-mirror system}

We shall first analyze the {\it stationary} resonances of the
charged black-hole-mirror system. In particular, in the present
section we shall derive an analytical formula for the {\it discrete}
radii of the mirror, $\{r^{\text{stat}}_{\text{m}}(\tau,qQ;n)\}$,
which satisfy the stationary resonance condition $\Im\omega=0$.
[Here $n=1,2,3,...$ is the resonance parameter]. It is worth noting
that the minimum radius of the mirror, $r_{\text{min}}\equiv
r^{\text{stat}}_{\text{m}}(\tau,qQ,n=1)$, would mark the boundary
between stable and unstable black-hole-mirror configurations:
configurations with $r_{\text{m}}<r_{\text{min}}$ would be stable
($\Im\omega<0$) whereas configurations with
$r_{\text{m}}>r_{\text{min}}$ would be unstable ($\Im\omega>0$).

The stationary resonances of the system (modes with $\Im\omega=0$)
are described by the field (\ref{Eq12}) with the critical frequency
for superradiance [see Eq. (\ref{Eq4})]:
\begin{equation}\label{Eq25}
\omega_{\text{c}}={{qQ}\over{r_+}}\  .
\end{equation}
The critical frequency (\ref{Eq25}) corresponds to the value
$\epsilon=0$ [see Eq. (\ref{Eq18})]. Taking cognizance of Eqs.
(\ref{Eq23})-(\ref{Eq24}) with $\epsilon=0$, one finds that the
`stationary' radial field $R^{\text{stat}}(x)$ \cite{Notestr} is
described by the Bessel function of the first kind (see Eq. 9.1.1 of
\cite{Abram}):
\begin{equation}\label{Eq26}
R^{\text{stat}}(x)=J_0\Big({{qQ}\over{\tau}}x\Big)\  .
\end{equation}

Taking cognizance of the boundary condition $R(x=x_{\text{m}})=0$,
which is dictated by the presence of the reflecting mirror [see Eq.
(\ref{Eq17})], one finds that the stationary resonances of the
charged field correspond to the discrete radii
\begin{equation}\label{Eq27}
x^{\text{stat}}_{\text{m}}(\tau,qQ;n)={{\tau}\over{qQ}}\times
j_{0,n}\ \ \ ; \ \ \ n=1,2,3,...
\end{equation}
of the mirror. Here $j_{0,n}$ is the $n$th positive zero of the
Bessel function $J_0(x)$. The real zeros of the Bessel functions
were studied by many authors \cite{Abram,Bes}. For completeness, we
state here the first three zeros of $J_0(x)$ \cite{Bes}:
$j_{0,1}=2.4048, j_{0,2}=5.5201$, and $j_{0,3}=8.6537$.

It is worth emphasizing that the smallest `stationary' radius of the
mirror, $x_{\text{min}}\equiv
x^{\text{stat}}_{\text{m}}(\tau,qQ;n=1)$, corresponds to the {\it
innermost} location of the mirror (for given values of the
parameters $qQ$ and $\tau$) which allows the extraction of the
Coulomb energy from the charged black hole. In other words, the
dimensionless radius $x_{\text{min}}$ marks the onset of instability
in the black-hole-mirror system: black-hole-mirror configurations
with $x_{\text{m}}<x_{\text{min}}$ are stable ($\Im\omega<0$)
whereas black-hole-mirror configurations with
$x_{\text{m}}>x_{\text{min}}$ are unstable ($\Im\omega>0$).

Note that the solution (\ref{Eq27}) with $qQ\gg1$ is consistent with
the near-horizon condition $x_{\text{m}}\ll\tau$ that we assumed
above [see Eq. (\ref{Eq22})]. In particular, one finds the
interesting property
\begin{equation}\label{Eq28}
x_{\text{min}}\to 0\ \ \ \text{as}\ \ \ qQ\to\infty\ .
\end{equation}
That is, the reflecting mirror can be placed arbitrarily close to
the black-hole horizon ($r_{\text{min}}\to r_+$) in the
$qQ\to\infty$ asymptotic limit.

\section{Rapidly growing superradiant instabilities}

The solution of the radial equation (\ref{Eq23}) obeying the ingoing
boundary condition (\ref{Eq16}) at the black-hole horizon is given
by \cite{Abram,Stro}
\begin{equation}\label{Eq29}
R=x^{-i\epsilon/\tau}{_2F_1}({1/2},{1/2}-2iqQ-4i\epsilon;1-2i\epsilon/\tau;-x/\tau)\
,
\end{equation}
where $_2F_1(a,b;c;z)$ is the hypergeometric function.

Our goal in the present section is to determine analytically the
resonance frequency $\omega=\omega(\tau,qQ,x_{\text{m}})$ which
satisfies the mirror-like boundary condition $R(x=x_{\text{m}})=0$
for a given radius $x_{\text{m}}$ of the mirror. It proofs useful to
use the ansatz \cite{Noteep}
\begin{equation}\label{Eq30}
\epsilon\equiv-qQx_{\text{m}}(1-\delta)\
\end{equation}
with $|\delta|\ll1$ [see Eq. (\ref{Eq40}) below], where the unknown
quantity $\delta=\delta(\tau,qQ,x_{\text{m}})$ is to be determined
below. This quantity contains within it the information about the
instability timescale (the value of $\Im\omega$) which characterizes
the composed black-hole-mirror system in the superradiant regime.

In the asymptotic regime \cite{Notexmin}
\begin{equation}\label{Eq31}
qQ\gg{{\tau}\over{x_{\text{m}}}}\gg1
%{{\tau}/{qQ}}\ll x_{\text{m}}\ll\tau
\end{equation}
one may use the large-$|b|$ asymptotic expansion \cite{Itn,Notef12}
\begin{eqnarray}\label{Eq32}
_2F_1(a,b;c;z)={{\Gamma(c)}\over{\Gamma(c-a)}}(-bz)^{-a}[1+O(|bz|^{-1})]\\
\nonumber
+{{\Gamma(c)}\over{\Gamma(a)}}(bz)^{a-c}(1-z)^{c-a-b}[1+O(|bz|^{-1})]
\end{eqnarray}
of the hypergeometric function in order to express the boundary
condition $R(x=x_{\text{m}})=0$ in the form
\begin{eqnarray}\label{Eq33}
i{{\Gamma(1/2)}\over{\Gamma(1/2-2i\epsilon/\tau)}}(1+{x_{\text{m}}/\tau})^{-2iqQ-4i\epsilon+2i\epsilon/\tau}
%e^{(1/2-2iqQ-4i\epsilon)x_{\text{m}}/\tau}
\\ \nonumber
\times[-(1/2-2iqQ-4i\epsilon)x_{\text{m}}/\tau]^{-2i\epsilon/\tau}&=&1\
.
\end{eqnarray}
The resonance condition (\ref{Eq33}) is a rather cumbersome equation
for the unknown quantity $\epsilon=\epsilon(\tau,qQ,x_{\text{m}})$
[or equivalently, for the unknown quantity
$\delta=\delta(\tau,qQ,x_{\text{m}})$]. Our goal is to solve the
resonance condition (\ref{Eq33}) analytically in the asymptotic
regime (\ref{Eq31}).

The inequality $qQx_{\text{m}}/\tau\gg1$ implies
$|\epsilon|/\tau\gg1$ [see Eqs. (\ref{Eq30}) and (\ref{Eq31})], in
which case one may use Eqs. 6.1.8 and 6.1.39 of \cite{Abram} to
write
\begin{equation}\label{Eq34}
{{\Gamma(1/2)}\over{\Gamma(1/2-2i\epsilon/\tau)}}\simeq{{(-2i\epsilon/e\tau)^{2i\epsilon/\tau}}
\over{\sqrt{2}}}\  .
\end{equation}
Substituting (\ref{Eq34}) into (\ref{Eq33}) one can write the
resonance condition in the form
\begin{eqnarray}\label{Eq35}
i2^{-1/2}(1+{x_{\text{m}}/\tau})^{-2iqQ-4i\epsilon+2i\epsilon/\tau}
%e^{(1/2-2iqQ-4i\epsilon)x_{\text{m}}/\tau}
\\ \nonumber
\times[(1/2-2iqQ-4i\epsilon)ex_{\text{m}}/2i\epsilon]^{-2i\epsilon/\tau}&=&1\
.
\end{eqnarray}
Taking the logarithm of both sides of Eq. (\ref{Eq35}), one finds
\cite{Notelnn,Noteln}
\begin{eqnarray}\label{Eq36}
\epsilon\Big[\ln\big(-{{qQx_{\text{m}}}\over{\epsilon}}\big)+1+
{{8\epsilon+i}\over{4qQ}}+x_{\text{m}}\big(2-{1\over\tau}\big)\Big]
\\ \nonumber
+qQx_{\text{m}}\big(1-{{x_{\text{m}}}\over{2\tau}}\big)-{1\over4}\tau(\pi+i\ln2)=0
\end{eqnarray}
for the resonance condition.

Substituting the ansatz (\ref{Eq30}) into Eq. (\ref{Eq36}), one
finds after some tedious algebra that the resonance condition
(\ref{Eq36}) can be expressed as a quadratic equation for the
dimensionless quantity $\delta=\delta(\tau,qQ,x_{\text{m}})$
\cite{Notedel2}:
\begin{eqnarray}\label{Eq37}
iqQx_{\text{m}}\delta^2-4iqQx^2_{\text{m}}\delta+i{{qQx^2_{\text{m}}}\over{\tau}}+{{x_{\text{m}}}\over{2}}
\\ \nonumber
-{{i\tau}\over{2}}(\pi+i\ln2)=0\ .
\end{eqnarray}
Using the facts that $|\delta|\gg x_{\text{m}}$ \cite{NoteK} and
$qQx_{\text{m}}/\tau\gg1$ [see Eq. (\ref{Eq31})], one can simplify
Eq. (\ref{Eq37}) to yield
\begin{eqnarray}\label{Eq38}
\delta^2\simeq
-{{x_{\text{m}}}\over{\tau}}+{{\tau}\over{2qQx_{\text{m}}}}(\pi+i\ln2)\
.
\end{eqnarray}

In the asymptotic regime
\begin{equation}\label{Eq39}
qQ\gg \Big({{\tau}\over{x_{\text{m}}}}\Big)^2\gg1\  ,
\end{equation}
one finds from (\ref{Eq38}) the simple solution \cite{Notedel}
\begin{equation}\label{Eq40}
\delta\simeq i\sqrt{{{x_{\text{m}}}\over{\tau}}}\ .
\end{equation}
Taking cognizance of Eqs. (\ref{Eq18}), (\ref{Eq30}), and
(\ref{Eq40}), we finally find
\begin{equation}\label{Eq41}
\Im\omega={{qQ}\over{r_+}}\sqrt{{{x^3_{\text{m}}}\over{\tau}}}\
\end{equation}
for the imaginary part of the resonance frequency \cite{Noterat}.

Note that $\Im\omega>0$ [see Eq. (\ref{Eq41})], which implies an
{\it instability} of the charged black-hole-mirror system. Moreover,
the simple linear scaling $\Im\omega\sim qQ$ found for the imaginary
part of the resonance frequency implies an instability timescale
$1/\Im\omega$ which can be made arbitrarily short in the
$qQ\to\infty$ limit.

%from the inequalities in (\ref{Eq}) one learns that
%\begin{equation}\label{Eq11}
%\tau\ll\Im(\omega r_+)\ll\sqrt{qQ}\times\tau\  .
%%\times {{\ln2}\over{4\sqrt{\pi}}}\  .
%\end{equation}

\section{Summary and discussion}

Motivated by the well-known phenomenon of superradiant instability
of a rotating Kerr black hole enclosed in a cavity, we have explored
here the analogous phenomenon of superradiant instability of a
charged Reissner-Nordstr\"om black hole enclosed in a reflecting
cavity. Imposing a mirror-like boundary condition on a charged
scalar field in the vicinity of the black-hole horizon,
%$(r_{\text{m}}-r_+)/r_+\ll1$
it was shown that the confined field grows exponentially over time
in the superradiant regime (\ref{Eq4}). In particular, we derived
analytic expressions for the oscillation frequency $\Re\omega$ [see
Eqs. (\ref{Eq18}) and (\ref{Eq30})] and the instability growth
timescale $1/\Im\omega$ [see Eq. (\ref{Eq41})] of the confined
charged field in the asymptotic regime $qQ\gg\tau/x_{\text{m}}\gg1$.

The instability timescale $T_{\text{ins}}$ which characterizes the
composed black-hole-mirror system in the asymptotic regime $qQ\gg
(\tau/x_{\text{m}})^2\gg1$ \cite{Noteinter} is given by
\begin{equation}\label{Eq42}
T_{\text{ins}}\equiv1/\Im\omega={{r_+
\sqrt{\tau/x^3_{\text{m}}}}\over{qQ}} .
\end{equation}
Remarkably, the simple scaling $T_{\text{ins}}\sim r_+/qQ$ found in
the asymptotic regime (\ref{Eq39}) implies that the instability
growth timescale of the confined superradiant modes can be made
arbitrarily short in the $qQ\to\infty$ limit (in particular,
$T_{\text{ins}}$ can be made much shorter than the dynamical
timescale set by the mass $M$ of the black hole).

It should be emphasized that the present {\it analytic} study is
restricted to the linear regime. As we have shown, the instability
(exponential growth) of the confined superradiant modes can be
revealed at this linear level. However, a fully non-linear {\it
numerical} simulation of the charged scalar field dynamics
\cite{HodPircha} is required in order to explore the end-state of
this superradiant instability. One possible stationary end-state of
the system may be described by the stationary resonances discussed
in Sec. III. It has also been suggested that the end-point of the
instability is attained after a violent bosenova explosion
\cite{Yos}.

It is well known that RN black holes undergo a Schwinger discharge
on very short timescales \cite{DamRuf}. Thus, the charged black hole
bomb probably has a limited astrophysical relevance. This charged
black-hole-mirror system should instead be regarded as a simple
toy-model for the astrophysically more relevant rotating
black-hole-bomb.
%This model may prove useful in future numerical
%studies aimed to investigate the (non-linear) end-state of the
%superradiant instabilities.

In this respect, the charged black-hole-mirror model has two
important advantages over the astrophysically more realistic
rotating black-hole bomb \cite{Dego}:
\newline
(1) Unlike the rotating Kerr black-hole spacetime which is not
spherically symmetric, the charged black-hole bomb can be ignited by
spherical modes. This spherical symmetry of the charged model is
expected to facilitate future non-linear numerical studies of the
superradiant instabilities.
\newline
(2) The unstable modes of the rotating black-hole bomb are
characterized by very small growth rates [see Eq. (\ref{Eq3})]. One
is therefore forced to use very long numerical integration times in
order to observe these weak superradiant instabilities [the
numerical integration time which is required in order to observe the
characteristic instabilities of the rotating black-hole-mirror
system should be of the order of $t_{\text{num}}\sim10^5M$, see Eq.
(\ref{Eq3})]. These extremely long integrations times may introduce
numerical errors into the system. On the other hand, we have seen
that the charged black-hole-mirror bomb is characterized by an
instability timescale (\ref{Eq42}) which can be made much shorter
than the instability timescale (\ref{Eq3}) of the rotating
black-hole-mirror bomb. Thus, moderate integration times would
probably be sufficient in order to explore the non-linear end-state
of the charged black-hole bomb.

These two important advantages of the charged black-hole-mirror bomb
make this system a convenient toy model for future numerical studies
aimed to investigate the non-linear dynamics of the explosive
superradiant instabilities.

\bigskip
\noindent
{\bf ACKNOWLEDGMENTS}
\bigskip

This research is supported by the Carmel Science Foundation. I thank
Yael Oren, Arbel M. Ongo and Ayelet B. Lata for helpful discussions.

%\newpage


\begin{thebibliography}{99}
\bibitem{Noteun} We use gravitational units in which $G=c=\hbar=1$.
\bibitem{Noteext} A maximally rotating (extremal) Kerr black hole
is characterized by $J=M^2$ with $M_{\text{ir}}=M/\sqrt{2}$.
\bibitem{Noteext2} A maximally charged (extremal) RN black hole
is characterized by $|Q|=M$ with $M_{\text{ir}}=M/2$.
\bibitem{Noteuni} Note that $q$ and $\mu$ stand for $q/\hbar$ and $\mu/\hbar$,
respectively. Thus, they have the dimensions of $($length$)^{-1}$.
\bibitem{NoteK} In obtaining Eq. (\ref{Eq24}) we have assumed that
$|(qQ+\epsilon)x^2|\ll|(qQ+2\epsilon)x+\epsilon|$ in Eq.
(\ref{Eq21}). In the $\epsilon=0$ case that we shall consider in
Sec. III below, this inequality is satisfied in the regime
$x_{\text{m}}\ll 1$ [see Eq. (\ref{Eq22})]. For the
$\epsilon=-qQx_{\text{m}}(1-\delta)$ case that we shall consider in
Sec. IV below, this inequality is satisfied in the regime
$x_{\text{m}}\ll|\delta|$. In addition, we have assumed that
$K^2/x(x+\tau)\gg\mu^2r^2_+(x+1)^2+l(l+1)$ in Eq. (\ref{Eq20}). In
the $\epsilon=0$ case that we shall consider in Sec. III below, this
inequality is satisfied in the regime
$(qQ)^2\gg{{{\tau}/{x_{\text{m}}}}}$ [see Eq. (\ref{Eq27}) below].
For the $\epsilon=-qQx_{\text{m}}(1-\delta)$ case that we shall
consider in Sec. IV below, this inequality is satisfied in the
regime $qQ\gg{{\tau}/{x_{\text{m}}}}$ [see Eq. (\ref{Eq31}) below].
\bibitem{Notestr} That is, the radial field $R(x)$ which corresponds to the stationary condition
$\Im\omega=0$.
\bibitem{Noteep} Note that we take $\Re\epsilon<0$ in the ansatz (\ref{Eq30}),
which corresponds to the superradiant regime (\ref{Eq4}) [see Eq.
(\ref{Eq18})].
\bibitem{Notexmin} Note that the regime (\ref{Eq31}) corresponds to
$x_{\text{min}}\ll x_{\text{m}}\ll \tau$ [see Eq. (\ref{Eq27})].
\bibitem{Notef12} The asymptotic expansion of $_2F_1(a,b;c;z)$, Eq. (\ref{Eq32}), is
valid in the regime $|b|\gg\text{max}\{|a|,|c|,|z|\}$ with $|bz|\gg
1$. The first inequality requires $qQ\gg |\epsilon|/\tau$, a
condition which is satisfied in the regime (\ref{Eq31}) [see Eq.
(\ref{Eq30})]. The second inequality requires
$qQx_{\text{m}}/\tau\gg1$, a condition which is also satisfied in
the regime (\ref{Eq31}).
\bibitem{Notelnn} Here we have used the relation $i=e^{i\pi(2n+{1\over2})}$, which implies
$\ln i=i\pi(2n+{1\over 2})$, where here $n=0,1,2,...$ is the
resonance parameter. We shall henceforth consider the fundamental
mode with $n=0$. In order to obtain the higher-order resonances one
should replace $\pi\to \pi(1+4n)$ in the equations below. The
imaginary part $\Im\omega$ of the resonance frequency is found to be
a decreasing function of the resonance parameter $n$ [see
\cite{Noteinter} below with $\pi\to \pi(1+4n)$].
\bibitem{Noteln} Here we have used the fact that $x_{\text{m}}/\tau\ll1$ [see Eq. (\ref{Eq31})], which implies
$\ln(1+x_{\text{m}}/\tau)\simeq
x_{\text{m}}/\tau-(x_{\text{m}}/\tau)^2/2$. In addition, we have
used the facts that $1/qQ\ll1$ and
$|\epsilon|/qQ=O(x_{\text{m}})\ll1$ [see Eq. (\ref{Eq30})], which
imply $\ln(1/2-2iqQ-4i\epsilon)\simeq\ln(-2iqQ)+(8\epsilon+i)/4qQ$.
\bibitem{Notedel2} Here we have used the ansatz (\ref{Eq30}) with
$|\delta|\ll1$, which implies
$\ln(-qQx_{\text{m}}/\epsilon)=-\ln(1-\delta)\simeq
\delta+\delta^2/2$. In addition, we neglected sub-leading terms
which are factors $O(x_{\text{m}})\ll1$ and $O(\delta)\ll1$ smaller
than the dominant terms which appear in Eq. (\ref{Eq37}).
\bibitem{Notedel} Note that the solution (\ref{Eq40}) for $\delta$ with $x_{\text{m}}/\tau\ll1$
[see Eq. (\ref{Eq31})] is consistent with the condition
$|\delta|\ll1$ that we assumed above. In addition, the requirement
$|\delta|\gg x_{\text{m}}$ \cite{NoteK} is satisfied thanks to the
inequalities $x_{\text{m}}\ll\tau\leq 1/\tau$.
\bibitem{Noteinter} In the intermediate asymptotic regime
$1\ll{{\tau}/{x_{\text{m}}}}\ll qQ\ll ({{\tau}/{x_{\text{m}}}})^2$
one finds from (\ref{Eq38}) the solution
$\delta\simeq\sqrt{(\pi+i\ln2)\tau/2qQx_{\text{m}}}$, which can be
approximated as
$\delta\simeq[1+i(2\pi)^{-1}\ln2]\sqrt{\pi\tau/2qQx_{\text{m}}}$.
Note that this solution with $\tau/qQx_{\text{m}}\ll1$ is consistent
with the condition $|\delta|\ll1$ that we assumed above. In
addition, the requirement $|\delta|\gg x_{\text{m}}$ \cite{NoteK} is
satisfied thanks to the inequalities $\tau/qQx_{\text{m}}\gg
x_{\text{m}}/\tau\geq x_{\text{m}}\gg x^2_{\text{m}}$. Taking
cognizance of Eqs. (\ref{Eq18}) and (\ref{Eq30}), one finds
$\Im(\omega r_+)\simeq \ln2\sqrt{\tau qQx_{\text{m}}/8\pi}$, which
implies $\Im\omega/\Re\omega=O(\sqrt{\tau x_{\text{m}}/qQ})\ll
x_{\text{m}}\ll1$ in this intermediate asymptotic regime. In this
regime one also finds $T_{\text{ins}}=1/\Im\omega=O(r_+/\sqrt{\tau
qQx_{\text{m}}})$ for the instability timescale.
\end{thebibliography}
\end{document}